\documentclass[%
reprint,
twocolumn,
superscriptaddress,
showpacs,preprintnumbers,
nofootinbib,
aps,
prd,
]{revtex4-2}

\usepackage{amsfonts}
\usepackage{bbm,bm}
\usepackage[breaklinks,urlbordercolor={1 1 1}]{hyperref}
\usepackage{graphicx}
\usepackage{amsmath}
\usepackage{amssymb}
\usepackage{mathtools}
\usepackage{hyperref}
\usepackage{graphicx}
\usepackage{color}
\usepackage{comment}
\usepackage{subfigure}
\usepackage{dcolumn}
\usepackage[utf8]{inputenc} 
\usepackage{multirow}
\usepackage{physics}

\usepackage{ulem,fancyvrb}
\usepackage{xcolor}

\begin{document}

\newcommand{\NOR}{\affiliation{Nordita,
KTH Royal Institute of Technology and Stockholm University, \\
Roslagstullsbacken 23,
SE-106 91 Stockholm,
Sweden }}

\newcommand{\TSUa}{\affiliation{Tsung-Dao Lee Institute \& School of Physics and Astronomy, Shanghai Jiao Tong University, Shanghai 200240, China }}

\newcommand{\TSUb}{\affiliation{Shanghai Key Laboratory for Particle Physics and Cosmology, Key Laboratory for Particle Astrophysics and Cosmology (MOE), Shanghai Jiao Tong University, Shanghai 200240, China }}

\newcommand{\AMH}{\affiliation{Amherst Center for Fundamental Interactions, Department of Physics,\\
University of Massachusetts, Amherst,
MA 01003, USA }}

\newcommand{\CAL}{\affiliation{Kellogg Radiation Laboratory, California Institute of Technology,\\
Pasadena,
CA 91125, USA}}
\newcommand{\PU}{\affiliation{Faculty of Fundamental Sciences, PHENIKAA University, Yen Nghia, Ha Dong, Hanoi 12116, Vietnam}}

\title{
Addressing the Gravitational Wave -- Collider Inverse Problem
}

\preprint{NORDITA 2022-010}

\author{Leon S.~Friedrich} \email{leonsteffenf@umass.edu} \TSUa \TSUb \AMH

\author{Michael J.~Ramsey-Musolf} \email{mjrm@sjtu.edu.cn, mjrm@physics.umass.edu} \TSUa \TSUb \AMH \CAL

\author{Tuomas V. I.~Tenkanen} \email{tuomas.tenkanen@su.se} \NOR \TSUa \TSUb

\author{Van Que Tran} \email{vqtran@sjtu.edu.cn; que.tranvan@phenikaa-uni.edu.vn} \TSUa \PU

\begin{abstract}

We provide a roadmap for analyzing the interplay between hypothetical future collider observations and the detection of a gravitational wave signal produced by a strong first order electroweak phase transition in beyond the Standard Model (BSM) theories. A cornerstone of this roadmap is a combination of a dimensionally reduced, three-dimensional effective field theory and results of both perturbation theory and non-perturbative lattice simulations. For the first time we apply these state-of-the-art methods to a comprehensive parameter space scan of a BSM theory. Concretely, we study an extension with the real scalar triplet, which admits a possible two-step electroweak symmetry-breaking thermal history. We find that 
(1) a first order transition during the second step could generate a signal accessible to LISA generation detectors and 
(2) the gravitational wave signal displays a strong sensitivity to the portal coupling between the new scalar and the Higgs boson, and
(3) the ability for future experiments to detect the produced gravitational waves depends decisively on the wall velocity of the bubbles produced during the phase transition. We illustrate how a combination of direct and indirect measurements of the new scalar properties, in combination with the presence or absence of a gravitational wave detection, could test the model and identify the values of the model parameters. 

\end{abstract}

\maketitle

In the Standard Model (SM) of particle physics electroweak symmetry  is smoothly  broken through a crossover transition as the temperature  is  lowered below the electroweak scale~\cite{Kajantie:1996mn,Csikor:1998eu}. However, the nature of the transition can be modified in the presence of physics beyond the Standard Model. Extended Higgs sectors in particular can lead to a first order electroweak phase transition (EWPT) with several associated theoretical and phenomenological implications. Such a phase transition could provide the necessary preconditions for the generation of the observed baryon asymmetry via  electroweak baryogenesis~\cite{Kuzmin:1985mm,Shaposhnikov:1987tw,Morrissey:2012db}.
Experimental signatures include deviations from SM predictions for various high energy collider observables~\cite{Ramsey-Musolf:2019lsf} -- at both present day and planned future next-generation detectors -- as well as production of primordial gravitational waves (GW) that may be accessible by next-generation experiments~\cite{Grojean:2006bp,Caprini:2019egz,LISACosmologyWorkingGroup:2022jok}. The latter include LISA \cite{Audley:2017drz}, DECi-hertz Interferometer GW Observatory (DECIGO)~\cite{Kudoh:2005as,Kawamura:2011zz}, Big Bang Observer (BBO)~\cite{Yagi:2011wg}, TAIJI~\cite{Gong:2014mca} and TIANQIN \cite{TianQin:2015yph}. These experiments would provide a looking glass into a cosmic era that significantly pre-dates the epoch of recombination. Thus, the combination of collider and GW probes hold the prospect of revealing in detail the BSM physics of the hot plasma consisting of particles and interactions at the time of electroweak symmetry breaking.

Many  BSM theories  can accommodate a first order EWPT -- which proceeds via bubble nucleation -- when the new degrees of freedom lie  close to the electroweak scale and couple with sufficient strength to the SM Higgs boson. The EWPT, thus, provides a clear target for collider and GW experiments~\cite{Ramsey-Musolf:2019lsf,Caprini:2019egz}. One then encounters two questions:
 
\noindent 1.
The \lq\lq GW -- collider inverse problem": can a combination of collider and GW observations be used to determine the BSM scenario responsible for the observed signals, and can these GW observables be used to measure relevant model parameters?

\noindent 2. Theoretical robustness: how reliable are the computations that attempt to address the first question?

Several studies addressing the GW-collider inverse problem have been published in the
past decade (c.f. \cite{Caprini:2015zlo,Caprini:2019egz} and e.g. \cite{Hashino:2016xoj,Beniwal:2017eik,Kang:2017mkl,Chala:2018opy,Bian:2019bsn,Ellis:2020awk}). 
Most have solely relied on the use of perturbation theory to analyze the EWPT thermodynamics and nucleation dynamics. However, enhanced thermal contributions from bosonic infrared modes render perturbation theory suspect in this context~\cite{Linde:1980ts}.

For quantitative and, in some cases, even qualitative reliability, it is critical to resort to non-perturbative approaches. Such non-perturbative studies are provided by lattice simulations
\cite{Farakos:1994xh,Kajantie:1995kf,Laine:2000rm,Laine:2012jy,Niemi:2020hto,Gould:2021dzl}
of dimensionally reduced \cite{Ginsparg:1980ef,Appelquist:1981vg},
three-dimensional effective field theories (3d EFT) \cite{Kajantie:1995dw,Braaten:1995cm}.%
\footnote{
Recent \cite{Ekstedt:2022bff} provides a public package for 3d EFT construction in generic models.
}
In this framework, recent work \cite{Gould:2019qek}
yields the model-independent conclusion that a GW signal cannot be accessible to LISA generation experiments, if the new scalar is sufficiently heavy that transition dynamics are described by the SM-like 3d EFT. Evidently, 
a GW signal detectable by LISA generation experiments
is likely only if new scalar is light enough to be dynamical and actively present in simulations. Such simulations were recently performed in \cite{Niemi:2020hto} 
for the real triplet extension of the SM ($\Sigma$SM), which non-perturbatively confirmed the possibility of two-step electroweak symmetry-breaking in the early universe that had been proposed by previous perturbative studies~\cite{Patel:2012pi,Blinov:2015sna,Inoue:2015pza}. 
See also similar recent study \cite{Niemi:2024axp} for the real singlet-extended SM.  

While a fully non-perturbative treatment of the $\Sigma$SM bubble dynamics (as in \cite{Moore:2000jw,Moore:2001vf,Gould:2022ran}) remains to be implemented, it is nevertheless of interest to address the GW-collider inverse problem (c.f.~\cite{Gowling:2021gcy}) by drawing on the 
cutting-edge perturbative treatment for
thermodynamics and results from non-perturbative (lattice) simulations.
Doing so is the goal of this letter.
We emphasize that use of perturbation theory is essential in order to draw conclusions regarding the full parameter space of a BSM theory; due to the large computational demand, non-perturbative studies are limited to selected benchmark points.
Our objective is to provide a template for future studies that may be performed for other models and, ultimately, rely on further advances in the theory of bubble dynamics. In addition to reliance upon the combination of lattice and EFT computations for a dynamical BSM scalar, the novel features of this work include:
\begin{itemize} 
\item We map the relevant portion of the model phase diagram into the plane of the key inputs for GW signals,  $(\alpha,\beta/H_*)$ \cite{Weir:2017wfa,Caprini:2019egz}, and
determine in detail how these inputs evolve with the parameters that characterise the phase diagram. 
To our knowledge, such a mapping and determination of this evolution has not yet been carried out in the full parameter space for any phenomenologically-viable BSM scenario while drawing upon state-of-the-art thermodynamics discussed above.
Importantly, our analysis combines
multiple developments in computation of the thermodynamics by utilizing the
3d EFTs, including non-perturbative information from lattice studies   
(c.f. \cite{Niemi:2018asa,Gould:2019qek,Niemi:2020hto,Croon:2020cgk,Gould:2021oba,Gould:2021ccf}).
\item We explore the relation between future collider physics  phenomenology and GW probes.  In doing so, the specific numbers we assume for hypothetical collider measurements 
are less important than the larger lesson that we illustrate:
how the combination of results can address the GW-collider inverse problem.
\end{itemize}

The choice of the $\Sigma$SM is particularly well-suited for the illustration of this theoretical \lq\lq template". It entails the minimum number of new scalar degrees of freedom that are charged under the SU(2$)_L$ electroweak gauge group; if a $Z_2$-symmetry is imposed, only a relatively small number of new parameters are introduced -- the triplet scalar mass ($m_\Sigma$), self-coupling  ($b_4$), and Higgs portal coupling ($a_2$) -- thereby enabling predictivity; and it allows for the non-trivial thermal history of multi-step electroweak symmetry-breaking.  The first transition, when it is first order, arises via a radiatively induced barrier. The second comes from a tree-level barrier in the presence of the thermal loop contributions to the potential. Even for a dynamical triplet scalar, a LISA-generation experimental signal is likely possible only from the second step in the two-step scenario.
We learn that
the value of the Higgs portal coupling  is decisive for this accessibility. 
Moreover, 
strength and duration of the transition depend strongly on $a_2$; $\mathcal{O}(1\%)$ changes in its value can lead to order of magnitude changes in the strength and duration.
Thus, the GW search provides a particularly powerful probe of the Higgs portal coupling.
In the collider arena, the Higgs diphoton decay rate, as well as the neutral triplet diboson decay rate,  have an analogous sensitivity to $a_2$ as well as sensitivities to the new scalar mass. Thus, a combination of GW and collider searches provide a means of identifying the relevant region of parameter space for a two-step transition that is unlikely to be obtainable by either probe alone.

We consider the most general renormalizable scalar potential for the $\Sigma$SM \cite{FileviezPerez:2008bj, Patel:2012pi, Niemi:2018asa,Bell:2020gug,Niemi:2020hto, Bell:2020hnr}
\begin{align}
\label{eq:V0} 
V(H,\Sigma) \ =& \  -\mu_h^2 H^\dagger H\  -\  \frac{1}{2}\mu_\Sigma^2
\mathrm{Tr}(\Sigma^2)\ \nonumber \\
&+\ \lambda_h (H^\dagger H)^2\ 
+\  \frac{1}{4} b_4 [\mathrm{Tr}(\Sigma^2)]^2  \ \nonumber \\
&+\ \frac{1}{\sqrt{2}} a_1 H^\dagger \Sigma H \ 
+ \ \frac{1}{2} a_2 \mathrm{Tr}(\Sigma^2) H^\dagger H \,, 
\end{align}
where $H$ is the SM Higgs doublet
and
the real scalar 
\begin{equation}
    \Sigma = 
    \begin{pmatrix}
        \Sigma^0/\sqrt{2} & \Sigma^+ \\
        \Sigma^- & -\Sigma^0/\sqrt{2}
    \end{pmatrix}
\end{equation}
transforms as $(1,3,0)$ under the SM $SU(3)_C\times SU(2)_L\times U(1)_Y$ gauge group
and is comprised of new neutral and charged particles $\Sigma^0$ and $\Sigma^\pm$, respectively. 
The triplets are primarily pair produced at colliders via charged- and neutral-current Drell-Yan processes $q \bar{q} \to \Sigma^+ \Sigma^-$ and $q \bar{q} \to \Sigma^{\pm} \Sigma^0$ \cite{FileviezPerez:2008bj}. 
For the non-$Z_2$ symmetric case, the neutral triplet has several interesting  decay channels,
such as decay into gauge boson pairs. 
Here, 
we focus on the decay $\Sigma^0 \to ZZ$ and additionally on 
$\Sigma^\pm$ loop-induced modifications of the Higgs to diphoton decay rate. 
Selected formulae for partial widths, branching fractions and decay rates are collected in the Appendix~\ref{sec:appendix-pheno}. These observables may be probed with future data at the LHC as well as prospective future colliders, including the Future Circular Collier (FCC) \cite{FCC:2018evy}, Circular Electron-Positron Collider (CEPC) \cite{CEPCStudyGroup:2018ghi}, International Linear Collider (ILC) \cite{Baer:2013cma}, Compact Linear Collider (CLIC) \cite{deBlas:2018mhx,Robson:2018zje}, or a high energy muon collider \cite{Palmer:2014nza, Delahaye:2019omf}. 

We employ thermal quantum field theory methods to obtain the quantities that govern the GW spectrum: 
$T_p$, the percolation temperature related to bubble nucleation; 
$\alpha$ at $T_p$,
which describes the strength of the transition in terms of the trace anomaly, related to entropy density and pressure between the phases;
and $\beta/H_*$, inverse duration of the transition, evaluated at $T_p$. For exact definitions, see \cite{Caprini:2019egz,Croon:2020cgk,Gowling:2021gcy}.
The thermodynamics of the 
EWPT in $\Sigma$SM 
were first studied in \cite{Patel:2012pi} 
using 4d perturbation theory and the $\hbar$-expansion to ensure gauge invariance \cite{Patel:2011th}.
As observed in that work, and subsequently verified in Refs.~\cite{Patel:2012pi,Blinov:2015sna,Inoue:2015pza,Niemi:2020hto}, the model  allows for a \lq\lq two-step\rq\rq\,  EWPT. During the first transition, wherein a thermal loop-induced barrier separates the symmetric and broken phases, the triplet acquires a non-zero 
vacuum expectation value (VEV). The second transition to the Higgs phase, which entails traversing a tree-level barrier, can be a sufficiently strong first order EWPT to provide for an observable GW signal. Here, we focus solely on these strong second transitions.

In our computation we work in 3d EFT with 
next-to-leading order
dimensional reduction \cite{Kajantie:1995dw,Niemi:2018asa}, and compute the effective potential \cite{Farakos:1994kx,Niemi:2020hto} and the bubble nucleation rate in analogy to analysis in \cite{Croon:2020cgk,Gould:2021oba} (c.f. \cite{Gould:2021ccf}), to obtain $T_p, \alpha, \beta/H_*$, described 
above. 
These parameters can be used in
{\tt PTplot} tool \cite{Caprini:2019egz} to obtain GW power spectrum $\Omega(f)$ as function of frequency, and LISA signal-to-noise ratio (SNR). 
The detailed calculations of the thermal parameters are shown in Appendix~\ref{sec:appendix-thermal}.
For relativistic hydrodynamic simulations of GW production from first order phase transitions, see  \cite{Hindmarsh:2013xza,Hindmarsh:2015qta,Hindmarsh:2017gnf,Cutting:2018tjt,Cutting:2019zws,Cutting:2020nla,Dahl:2021wyk}.

The regions of the two-step viable parameter space have been identified using a combination of the 3d EFT perturbation theory and lattice 
simulations
involving a dynamical triplet in Ref.~\cite{Niemi:2020hto}.
In analogy to \cite{Gould:2021oba}, we determine bubble nucleation rate, from which we obtain $T_p$ and $\beta/H_*$, only at leading order in 3d EFT, which acts as the limiting theoretical uncertainty in our computation \cite{Croon:2020cgk,Gould:2021oba}. 
In particular, in terms of the EFT description of \cite{Gould:2021ccf} -- where nucleating fields represent zero Matsubara modes of the original theory before dimensional reduction  -- we compute the leading order effective action within the EFT, but emphasize that several higher order, hard thermal scale contributions are resummed therein, such as two-loop thermal masses. For consistency, we also use leading order result for $\alpha$. 

\begin{figure}
   \centering
   \includegraphics[width=0.45\textwidth]{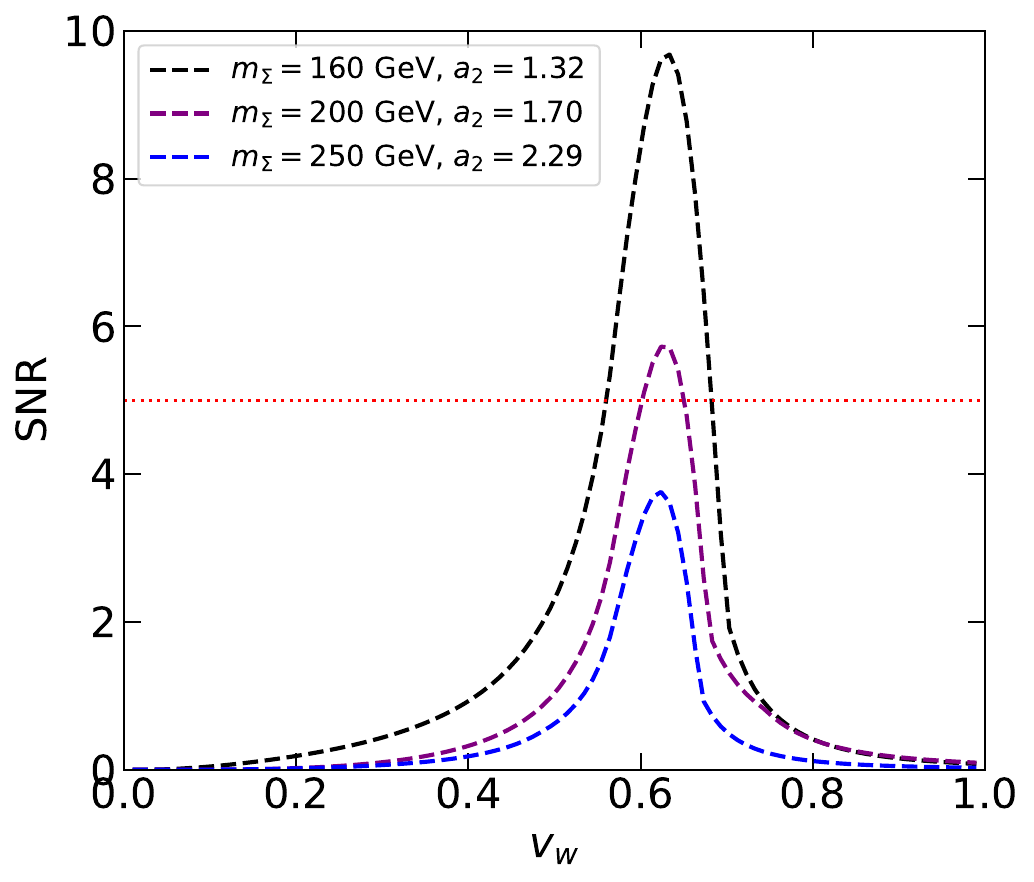}
   \caption{
    The LISA SNR as a function of wall velocity. The dashed black, purple, and blue lines correspond to benchmark points with ($m_\Sigma$, $a_2$, $b_4$) values of (160 GeV, 1.32, 0.1), (200 GeV, 1.70, 0.1), and (250 GeV, 2.29, 0.1), respectively. The dotted red line indicates ${\rm SNR} = 5$.}
    \label{fig:vw}
\end{figure}

We consider the wall velocity ($v_w$) as a crucial input parameter for gravitational wave (GW) production, yet its exact value remains unknown. For our analysis, we set $v_w = 0.63$, chosen to optimize the SNR prediction at LISA, as demonstrated in Fig.~\ref{fig:vw}.
For recent development on computing $v_w$ from first principles, see \cite{Bodeker:2017cim,Hoche:2020ysm,Baldes:2020kam,Azatov:2020ufh,Gouttenoire:2021kjv,Laurent:2022jrs,Krajewski:2024gma}.

\begin{figure*}[t]
   \centering
   \includegraphics[width=0.45\textwidth]{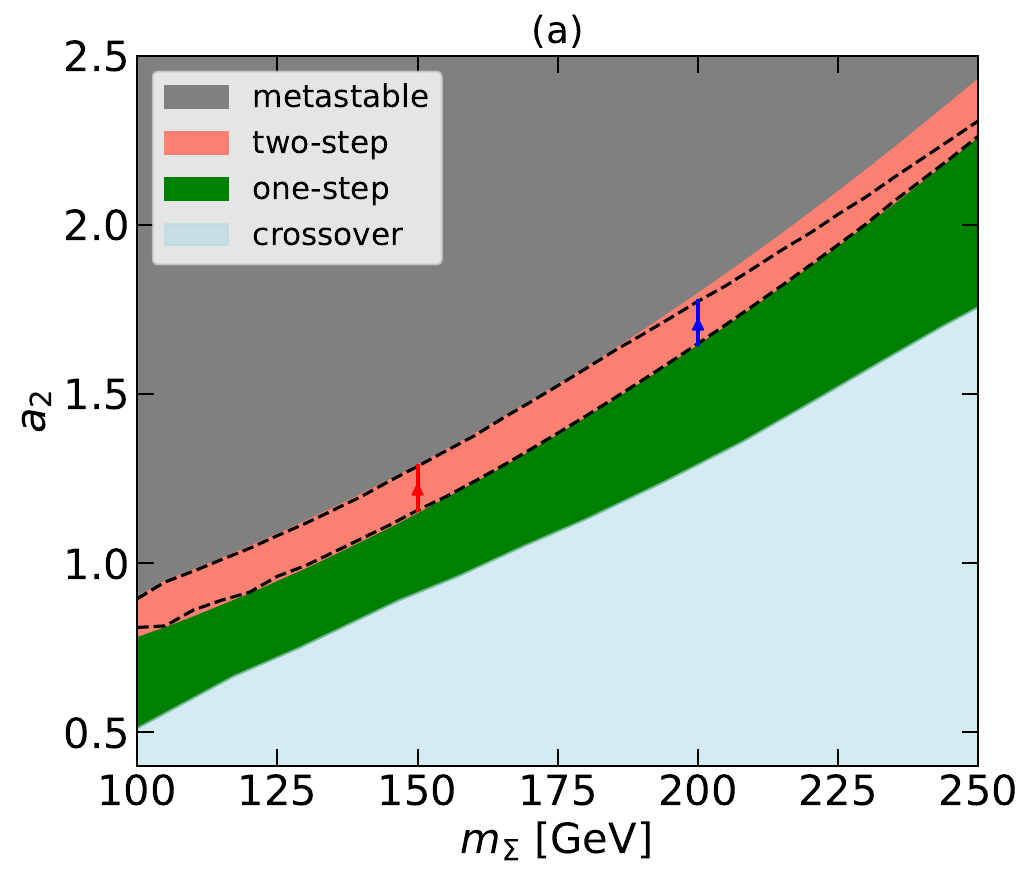}
   \includegraphics[width=0.45\textwidth]{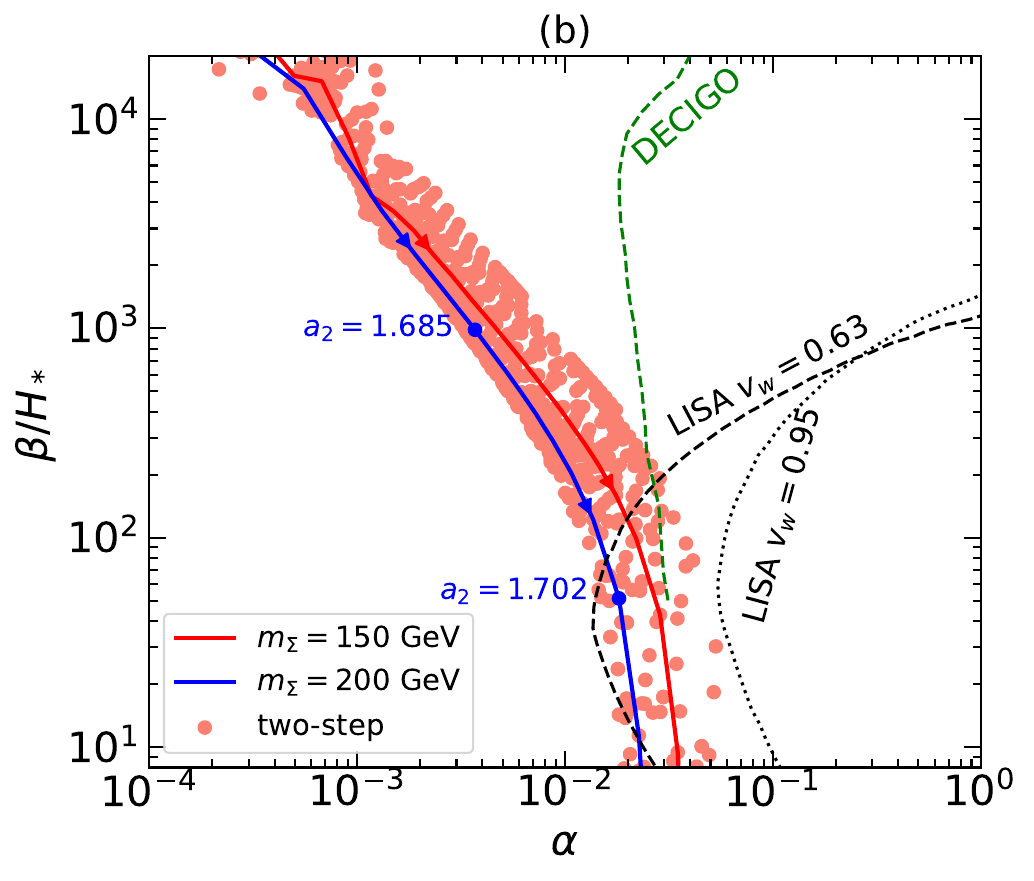}
   \includegraphics[width=0.45\textwidth]{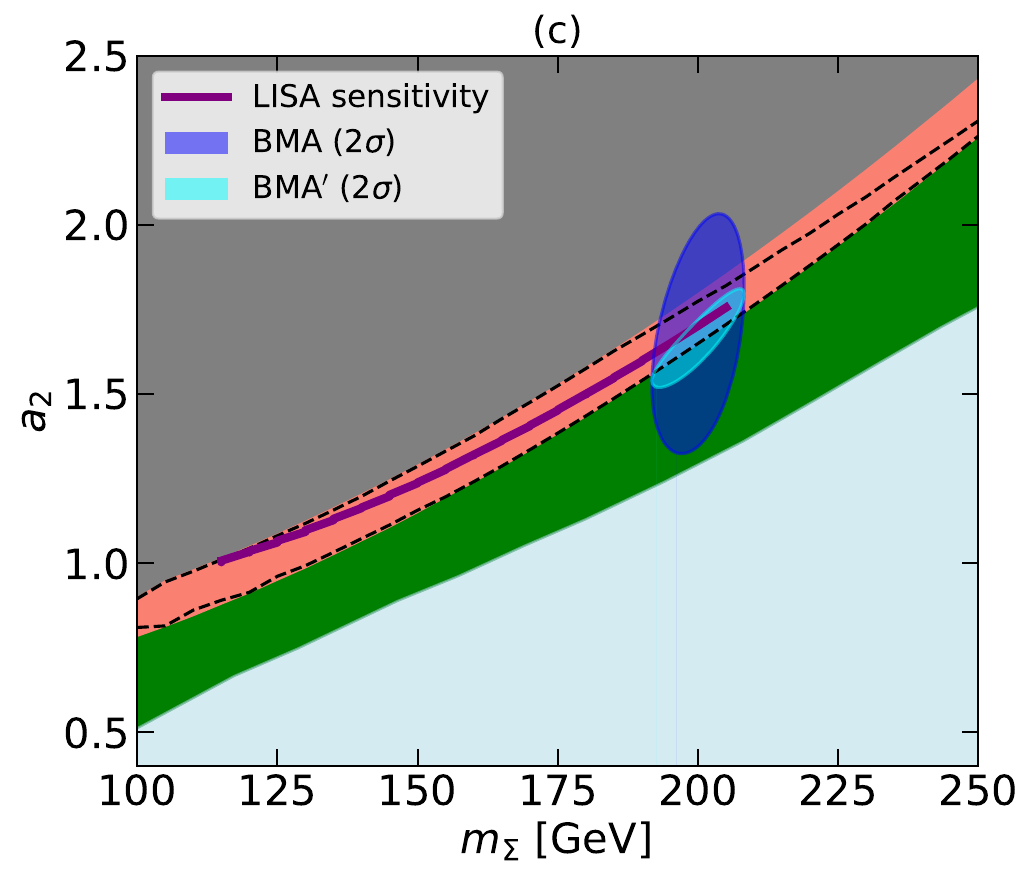}
   \includegraphics[width=0.45\textwidth]{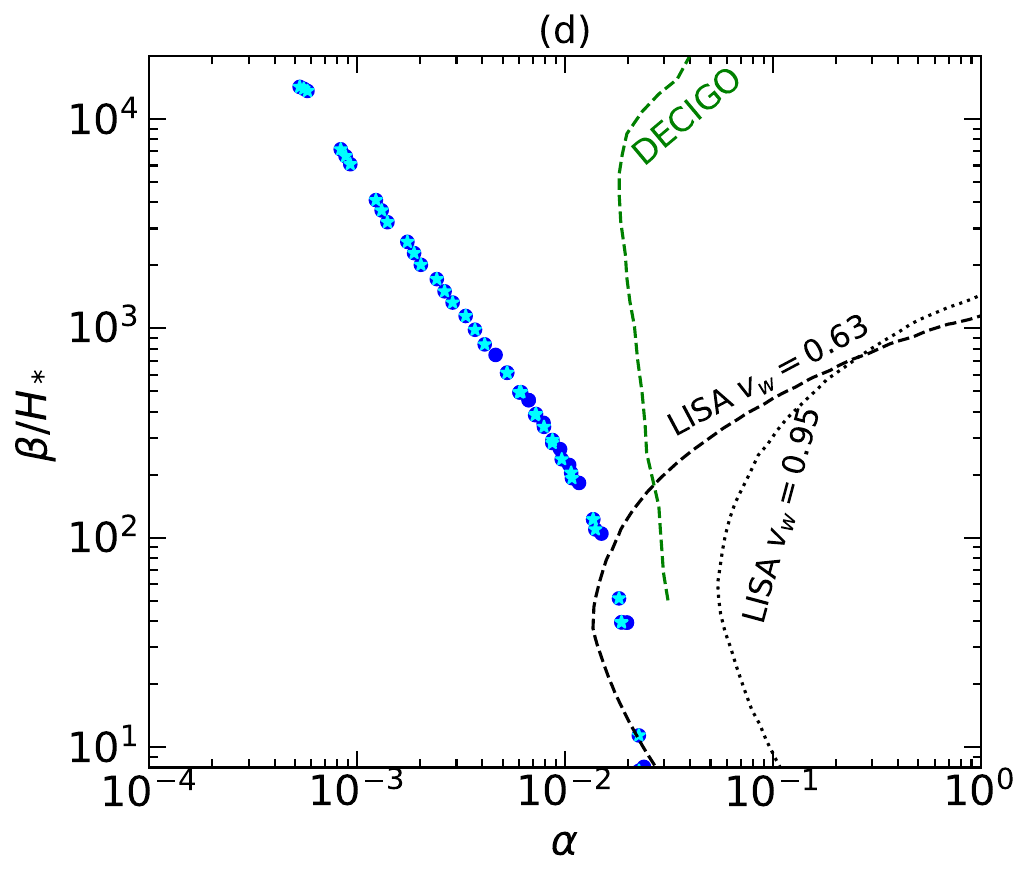}
   \caption{
    Panel (a): The phase structure of the real-triplet model is depicted in the ($m_{\Sigma}, a_2$) plane, with $b_4$ fixed at $1.0$. The grey, pink, green, and light blue regions represent the metastable electroweak minimum, two-step, one-step first order phase transition, and crossover transition, respectively. Inside the dashed black contour within the two-step region, bubble nucleation completes.
    Panel (b): Results for completed two-step transitions are displayed in the ($\alpha, \beta/H_{*}$) plane as a scatter plot of light pink points. Solid red and blue lines represent results with fixed $m_\Sigma = 150$ GeV and $200$ GeV, respectively, as also depicted in panel (a); arrows indicate the increase of the coupling $a_2$. Additionally, dashed green, dotted, and dashed black lines indicate sensitivities from DECIGO with fixed $v_w = 0.95$, LISA with fixed $v_w = 0.95$, and LISA with fixed $v_w = 0.63$, respectively. 
    Panel (c): The $2\sigma$ favored regions from hypothetical collider measurements BMA (BMA$'$) are depicted as blue (cyan) regions, and the sensitivity from LISA with SNR $\geq 5$ is shown as a purple band, projected onto the phase structure.
    Panel (d): The $2\sigma$ favored points of BMA (blue circle) and BMA$'$ (cyan star) are plotted in the ($\alpha, \beta/H_{*}$) plane.}
    \label{fig:fig1}
\end{figure*}

Our numerical analysis consists of 
two parts: 

\noindent 1.
We scan through the whole model parameter space $(m_\Sigma,a_2,b_4)$ to identify the two-step phase transition regions \cite{Niemi:2020hto}.%
\footnote{
In the non-$Z_2$ symmetric 
case,
there is also non-vanishing $a_1$ parameter. However, it is constrained to be small and has negligible effect for thermodynamics \cite{FileviezPerez:2008bj,Patel:2012pi}.
}
For each point therein we compute the GW parameters, 
$\Omega(f)$, and LISA SNR value. Assuming an observation occurs for $\text{SNR} \geq 5$ \cite{Smith:2019wny}, we project the result back onto the ($a_2$, $m_\Sigma$) plane.

\noindent 2.
We provide a template 
for future collider diagnostics, 
by selecting a benchmark parameter point inside the two-step region and 
by envisioning hypothetical measurements. 
We then identify the hypothetical collider-allowed region in the 
($a_2$, $m_\Sigma$) and ($\alpha$, $\beta/H_\ast$) planes.

We start from 1:
Fig.~\ref{fig:fig1}(a) shows 
the 
model
phase structure projected in the ($m_\Sigma, a_2$) plane with 
fixed coupling $b_4 = 1.0$.
The pink region indicates the parameters accommodating a two-step EWPT history. Therein, the nucleation process does not complete for parameters  outside the black dashed contour. The green and light blue areas correspond to a one-step first order EWPT and crossover transition, respectively.  We use non-perturbative information from lattice simulations of \cite{Kajantie:1995kf} along lines of \cite{Brauner:2016fla,Andersen:2017ika,Niemi:2018asa,Gould:2019qek} in order to delineate the boundary between these two regions. 
Varying $b_4$ leads to mild changes in the EWPT-viable regions as detailed in Appendix~\ref{sec:appendix-pheno}. 

Fig.~\ref{fig:fig1}(b) illustrates a scatter plot of the values of $\alpha$ and $\beta/H_{*}$ corresponding to the area enclosed by the black dashed curve in Fig.~\ref{fig:fig1}(a). 
The dashed green line indicates the experimental sensitivity of DECIGO with the ``Correlation" design \cite{Kawamura:2011zz}, with $v_w$ fixed at 0.95. The dotted and dashed black lines represent the sensitivity threshold of SNR=5 for LISA, with $v_w$ fixed at 0.95 and 0.63 respectively \cite{Caprini:2019egz}. The region probed by LISA is significantly influenced by the chosen value of $v_w$. Opting for $v_w = 0.63$ enables the exploration of a portion of the two-step region, whereas none of it is accessible with $v_w = 0.95$. However, DECIGO with $v_w = 0.95$ has the capability to probe a part of the two-step region.
\footnote{
For illustration of this line, we have fixed  $T_* = 100$ GeV, while for all scatter points we have specific $T_p$ determined individually for each point.
}

It is particularly interesting to determine the dependence of the GW parameters on $a_2$ and $m_\Sigma$. To that end, for visualisation we choose two values of the latter, $m_\Sigma=$ 150 and 200 GeV, corresponding to the red and blue lines, respectively, in
Fig.~\ref{fig:fig1}(a).
Arrows indicate increasing $a_2$. 
We recast these lines in 
Fig.~\ref{fig:fig1}(b),
thereby depicting the corresponding trends in $\alpha$ and $\beta/H_\ast$. 
It is 
evident that the GW parameters carry a pronounced sensitivity to the Higgs portal coupling and a milder dependence on 
the triplet scalar mass. In particular,  a larger $a_2$ results in a stronger phase transition and a larger transition duration time. 
For instance, for the case $m_\Sigma = 200$ GeV, at $a_2 = 1.685$
the values of $\alpha \sim 3\times 10^{-3}$ and $\beta/H_{*} \sim 10^3$, but increasing $a_2$ by $\sim1\%$ results in $\alpha$ ($\beta/H_{*}$) increasing (decreasing) by more than one order of magnitude. 
Moreover, a decrease in the mass of the triplet scalar leads to a stronger
phase transition, consequently shifting towards regions that will be probed by future GW detectors.

To illustrate the interplay with collider phenomenology and provide a template for future analysis (point 2 above)
we 
envision a hypothetical set of measurements:
\begin{align}
\delta_{\gamma\gamma} &= -0.132 \pm 0.015, \\  
m_\Sigma &= (200 \pm 5 ) \; \text{GeV} \\
\text{BR}(\Sigma^0 \to ZZ) &= 0.01 \pm 0.002 \ \ \ ,
\end{align}
where $\delta_{\gamma\gamma}$ gives the change of the Higgs diphoton decay rate relative to the SM prediction and where the decay $\Sigma^0\to ZZ$ occurs only for the non-$Z_2$ symmetric version of the model ($a_1\not=0$). For these two quantities, 
the central values correspond to
$a_2 = 1.665$ and $b_4 = 1.0$.
The uncertainty on the diphoton rate measurement is chosen to correspond to the sensitivity of future lepton collider FCC-ee \cite{dEnterria:2016fpc}. 
For a similar discussion in case of the measurements for HL-LHC, see Appendix~\ref{sec:appendix-pheno}.
For uncertainty in the triplet scalar mass, we pick an arbitrary, yet conservative estimate. 
The mass can be identified via a reconstruction of the 4 leptons (or $ZZ$) invariant mass if an excess of $ZZ$ pairs is observed. 
Note that if the triplet scalar is stable (due to $Z_2$ symmetry), its mass can be measured using the disappearing charged tracks search at the LHC \cite{CMS:2016kce, ATLAS:2017oal}.
The uncertainty in the $\text{BR}(\Sigma^0 \to ZZ)$ measurement 
is taken to be $20\%$ of the central value. 
We note that this choice for uncertainty is arbitrary, and used here solely for the purpose of illustration, indicating roughly the level of accuracy required to decisively indicate two-step phase transition.
We denote benchmarks without (with) $\text{BR}(\Sigma^0 \to ZZ)$ measurement as  BMA (BMA$'$).
In the Appendix~\ref{sec:appendix-pheno} we discuss in more detail how $\text{BR}(\Sigma^0 \to ZZ)$ and $\delta_{\gamma\gamma}$ depend on $m_\Sigma$ and $a_2$, and present another illustrative choices for benchmark points. 

The 2$\sigma$ favored region from the BMA (BMA$'$) measurement is represented by the blue (cyan) area in the ($m_\Sigma, a_2$) plane of Fig.~\ref{fig:fig1}(c).
We see  that BMA is consistent with either a first order transition or an unstable electroweak vacuum. For the cosmologically viable region, however, this combination of collider measurements is not by itself sufficient to discriminate between a one-step or two-step transition scenario. For the non-$Z_2$ symmetric version of the model, additional collider information from $\text{BR}(\Sigma^0 \to ZZ)$, as in BMA$'$, could indicate that the $2\sigma$ region lies mainly within two-step parameter space.

A GW signal could provide a complementary indicator, independent of the presence or absence of $Z_2$ symmetry. To illustrate, we recast the LISA sensitivity region from Fig.~\ref{fig:fig1}(b) (with $v_w = 0.63$) 
to Fig.~\ref{fig:fig1}(c) by narrow purple region.
In this example, the combination of GW observation and collider significantly narrows down the parameter region of BMA, as well as BMA$'$. This feature results,
in part, from the strong sensitivity of the GW signal to the Higgs portal coupling $a_2$.
It is also worth noting that the combination of collider and GW results provide an important consistency test of the model. For example, if the BMA$'$ region resulting from addition of the $\Sigma^0\to ZZ$ decay were not to overlap with the LISA band, then one would conclude that the GW signal results from a different source.

It is also interesting to ask how the collider discovery of a given model and constraints on its parameters might interface with a LISA null result. To illustrate this possible scenario, we project the BMA and BMA$'$ allowed points into the ($\alpha$, $\beta/H_\ast$) plane as depicted in
Fig.~\ref{fig:fig1}(d).
A negative result from LISA would consequently disfavour the points represented by the blue circle (BMA) and cyan star (BMA$'$) located within the region where $\text{SNR} \geq 5$. 
The remaining portion of the collider-allowed two-step region could, nevertheless, accommodate a future generation GW probes. Even in this case, however, development of even more sensitive GW probes would be needed to cover the entire collider-favoured, two-step parameter space. Only in that case would one be able to definitively conclude that the electroweak symmetry-breaking transition in this model was not first order (two-step). 

We conclude that strong two-step transitions in the $\Sigma$SM 
are highly sensitive to the triplet-Higgs portal coupling. Therefore, a detection of a GW background signal could lead to a precise determination of the portal coupling, providing a complementary probe to future high energy colliders. We expect this feature persists in the case of other scenarios. More generally, 
the foregoing discussion provides a roadmap for future studies of the GW-Collider inverse problem, whether in the $\Sigma$SM or other models. Key ingredients include (1) several \lq\lq upgrades\rq\rq in the use of perturbation theory treatment of EWPT thermodynamics and nucleation dynamics:%
\footnote{
Our cutting-edge perturbative computation of thermodynamics could still be improved. This is to go even further beyond by adding several higher order corrections in analogy to individual state-of-art computations presented in \cite{Niemi:2020hto,Croon:2020cgk,Ekstedt:2021kyx,Ekstedt:2022tqk,Giese:2020rtr,Giese:2020znk,Tenkanen:2022tly}. Also see \cite{Ekstedt:2023sqc}.  }
we include several higher order thermal resummations \cite{Niemi:2018asa,Niemi:2020hto};
our perturbative expansion is consistent in powers of couplings and includes renormalization group improvement related to hard thermal scale \cite{Arnold:1992rz,Ekstedt:2020abj,Gould:2021oba,Schicho:2022wty}; our analysis is properly gauge invariant \cite{Patel:2011th,Croon:2020cgk};
and we perform self-consistent computation of the bubble nucleation rate \cite{Croon:2020cgk,Gould:2021oba,Gould:2021ccf}. (2) The presence of the two-step EWPT-viable region of parameter space
and the character of the each transition has been determined by lattice simulations in \cite{Niemi:2020hto}. Findings therein support perturbative computations here. 
Future refinements should include a
non-perturbative determination of the bubble nucleation rate \cite{Moore:2000jw,Moore:2001vf} (also cf. recent \cite{Gould:2022ran}).
Perturbative 3d EFT computations here pave the way for such future lattice simulations, that would push the current state-of-the-art even further.
In addition, a computation of the bubble wall speed as function of model parameters \cite{Laurent:2022jrs} should still be included, as well as an improved computation of the GW spectrum from thermal parameters beyond approximations made in {\tt PTplot} \cite{Caprini:2019egz}. Indeed, our study illustrates the decisive impact of $v_w$ on the sensitivity of future GW detectors to an EWPT-generated signal. Finally, future work could include a statistical analysis answering to what degree of accuracy thermal parameters and underlying model parameters can be reconstructed from the GW spectra, c.f. e.g. \cite{Gowling:2021gcy,Boileau:2022ter,Gowling:2022pzb,Lewicki:2024xan,Caprini:2024hue}.

\begin{acknowledgments}

We thank Daniel Cutting, Andreas Ekstedt, Oliver Gould, Yann Gouttenoire, Chloe Gowling, Mark Hindmarsh, Joonas Hirvonen, Deanna Hooper, Benoit Laurent, Marek Lewicki, Johan L{\"o}fgren, Lauri Niemi, Philipp Schicho and Jorinde van de Vis
for illuminating discussions. Specifically we thank Oliver Gould for sharing a code for bubble nucleation in 3d EFT \cite{Gould:2021oba};
and David J. Weir for permission to use {\tt PTplot} \cite{Caprini:2019egz}.
TT thanks Aleksi Vuorinen and Helsinki Institute of Physics  -- 
and VQT would like to thank the Institute of Physics, Academia Sinica, Taiwan -- for their  hospitality during completion of part of this work.
This work has been supported in part under National Science Foundation of China grant no. 19Z103010239.

\end{acknowledgments}

\allowdisplaybreaks
\bibliographystyle{apsrev4-1}
\bibliography{references}

\onecolumngrid
\appendix
\allowdisplaybreaks

\section{Formulae for thermal parameters}
\label{sec:appendix-thermal}

Let us discuss in detail our computation of thermal parameters 
$T_*$, $\alpha$ and $\beta/H_*$ evaluated at $T_*$. Here $T_*$ is the temperature at which gravitational waves are generated. 
Parameters of the 3d EFT at high temperature are found by dimensional reduction at NLO, as described in \cite{Niemi:2020hto,Niemi:2018asa}. 
Renormalization scale of the 3d EFT is fixed to $\mu_3 = T$. Leading order effective potential of the 3d EFT reads
\begin{align}
\label{eq:Veff3d}
V^{{\rm LO}}_{{\rm eff}}(\phi_3, x_3) = \frac{1}{2} \mu^2_{h,3} \phi^2_3 + \frac{1}{2} \mu^2_{\Sigma,3} x^2_3 + \frac{1}{4} \lambda_{h,3} \phi^4_3 + \frac{1}{4} b_{4,3} x^2_3 + \frac{1}{4} a_{2,3} \phi^2_3 x^2_3,
\end{align} 
where the background fields $\phi_3, x_3$ have dimension $T^{\frac{1}{2}}$ and couplings have dimension $T$. We emphasize that all 3d EFT parameters include thermal resummations from non-zero Matsubara modes at NLO, in particular thermal masses are determined at two-loop order.
To guarantee the gauge invariance of the calculation, 
one can follow \cite{Croon:2020cgk,Niemi:2020hto,Gould:2021oba} and
expand the effective potential 
order-by-order in the loop-counting parameter $\hbar$. 
The expansion of the potential and its minima read formally
\begin{align}
\label{eq:Veff-expand}
V_{\rm eff}^{\hbar} = V_0 + \hbar V_1 + \hbar^2 V_2, \quad
    v_\text{min} = v_0 + \hbar v_1 + \hbar^2 v_2, \quad
    x_\text{min} = x_0 + \hbar x_1 + \hbar^2 x_2,
\end{align}
where $V_1, V_2$ are one- and two-loop corrections, respectively \cite{Niemi:2020hto}. 
By evaluating the effective potential at its minima, we obtain
\begin{align}
V_{\rm eff}^{\hbar}(v_\text{min}, x_\text{min}) =& V_0(v_0, x_0) + \hbar V_1(v_0, x_0) + \hbar^2 \left[V_2(v_0, x_0) - \frac12 v_1^2 \pdv[2]{V_0}{v} - \frac12 x_1^2 \pdv[2]{V_0}{x} - v_1 x_1 \pdv{V_0}{v}{x} \right] + \mathcal{O}(\hbar^3), 
\end{align}
where ${\cal O}(\hbar)$  corrections for the minima are given as
\begin{align}
\label{eq:v1x1}
v_1 =& \left[\left( \pdv{V_0}{v}{x} \right)^2 - \left(\pdv[2]{V_0}{v}\right) \left(\pdv[2]{V_0}{x} \right) \right]^{-1} \left[ \left( \pdv[2]{V_0}{x} \right)\left( \pdv{V_1}{v} \right) - \left( \pdv{V_0}{v}{x} \right)\left( \pdv{V_1}{x} \right) \right], \\
x_1 =& \left[\left( \pdv{V_0}{v}{x} \right)^2 - \left(\pdv[2]{V_0}{v}\right) \left(\pdv[2]{V_0}{x} \right) \right]^{-1} \left[ \left( \pdv[2]{V_0}{v} \right)\left( \pdv{V_1}{x} \right) - \left( \pdv{V_0}{v}{x} \right)\left( \pdv{V_1}{v} \right) \right],
\end{align}
where all derivatives are evaluated at the tree-level minima $(v_0, x_0)$.
To study the phase structure, we determine the evolution of $V_{\rm eff}^{\hbar}$ as a function of temperature with different phases. The critical temperature $T_c$ can be determined by solving the degeneracy condition of $V_{\rm eff}^{\hbar}$ in any two minima, \textit{e.g.} $V_{\rm eff}^{\hbar}[(0, x_{\rm min}), T_c] = V_{\rm eff}^{\hbar}[(v_{\rm min}, 0), T_c]$ for $\Sigma\rightarrow H$ transition. 

For thermal bubble nucleation rate, the leading approximation reads
\begin{align}
\Gamma = A(T) e^{- S_{\rm eff}^{\rm LO}(\Phi_{3})},
\end{align} 
where the prefactor is simply estimated by $A(T) = T^4$. In 3d EFT, the leading order action reads
\begin{align}
\label{eq:action}
S_{\rm{eff}}^{\rm LO}(\Phi_{3}) &= \int_{0}^{\infty} dr r^2 \frac{1}{2} \left( \pdv{\Phi_3}{r} \right)^2 + V^{\rm LO}_{\rm{eff}}(\Phi_3, T),
\end{align} 
where $\Phi_{3} = \{\phi_3, x_3\}$ denotes a collection of the background fields in 3d EFT, that minimizes the action $\mathcal S_{\rm{eff}}^{\rm LO}$ and are determined by solving the following equation of motion, 
\begin{equation}
\label{eq:bounceeq}
\frac{{\rm d}^2 \Phi_{3}}{{\rm d} r^2}+\frac{2}{r} \frac{{\rm d} \Phi_{3} }{{\rm d} r}=\frac{{\rm d} V^{\rm LO}_{\rm{eff}}(\Phi_{3}, T)}{{\rm d} r} \,, 
\end{equation}
with the boundary conditions
\begin{equation}
\lim_{r \rightarrow \infty} \Phi_3(r) = 0,\left.\quad \frac{{\rm d} \Phi_3}{{\rm d} r}\right|_{r=0}=0 \,. 
\end{equation}
We utilize the code {\tt FindBounce}~\cite{Guada:2020xnz} to numerically solve the bounce equation in Eq.~(\ref{eq:bounceeq}) and then evaluate the action in Eq. \ref{eq:action}.
The inverse duration of the phase transition can be determined from
\begin{equation}
\frac{\beta}{H_*} = {T_*}{\frac{d S^{\rm LO}_{\rm eff}}{dT}} |_{T_*}. 
\end{equation} 
The Hubble rate $H_*$ is assumed to correspond to the radiation dominated universe and is given as
\begin{align}
H^2_* = \frac{\rho_{\rm{rad}}}{3 M^2_{\rm Pl}},
\end{align}
with the reduced Planck mass of $M_{\rm Pl} = 1.220910 \times 10^{19}/\sqrt{8\pi}$ GeV and the radiation energy density
\begin{align}
\rho_{\rm rad} &= \frac{\pi^2}{30} g_* T^4. 
\end{align}
Here the number of relativistic degrees of freedom $g_* = 106.75+3$, where 3 triplet degrees of freedom are added to the SM value.
In this study, we assume that $T_* = T_p$, and follow the analysis in \cite{Enqvist:1991xw} to determine the percolation temperature, $T_p$, that is the temperature at which satisfies the condition $h(t_p) = 1/e$ where $h(t)$ is the fraction of space at time $t$. 
Solving this percolation condition leads to \cite{Enqvist:1991xw, Caprini:2019egz} 
\begin{align}
S(T_p)  \simeq  131 + \log(\frac{A}{T_p^4}) - 4 \log(\frac{T_p}{100 {\rm GeV}}) - 4 \log(\frac{\beta/H}{100}) + 3 \log(v_w). 
\end{align}
We fix the bubble wall speed $v_w = 0.63$. 
The phase transition strength (the trace of the energy momentum weighted by the enthalpy) can be given as%
\footnote{
Note that $V_{\rm eff}^{\hbar}$ lives in the 3d EFT, and is related to the effective potential of parent 4d theory by $V^{\rm 4d}_{\rm eff} \sim T V_{\rm eff}^{\hbar}$.
} 
\begin{align}
\alpha = \frac{T}{\rho_{\rm{rad}}} \Big( T \frac{d \Delta V_{\rm eff}^{\hbar}}{dT} - 3 \Delta V_{\rm eff}^{\hbar}\Big)/4 , 
\end{align}
where $\Delta V_{\rm eff}^{\hbar}$ is the difference in effective potential between the lower and higher phases. Since we do not include higher order soft corrections, i.e. corrections within the EFT, to the bubble nucleation rate, for consistency we also determine $\alpha$ at leading order within the EFT.

Finally, we employ the {\tt PTPlot} package \cite{Hindmarsh:2017gnf, Caprini:2019egz} to compute the GW spectrum. To assess the detectability of the signals, one can define the signal-to-noise ratio (SNR) \cite{Caprini:2019egz} as follows
\begin{equation}
\mathrm{SNR}=\sqrt{\mathcal{T} \int_{f_{\min }}^{f_{\max }} \mathrm{d} f\left[\frac{h^2 \Omega_{\mathrm{GW}}(f)}{h^2 \Omega_{\mathrm{exp}}(f)}\right]^2},
\end{equation}
where $\mathcal{T}$ represents the duration of the observation period in years, $h^2 \Omega_{\mathrm{GW}}(f)$ denotes the spectrum of the fraction of GW energy from the FOEWPT, and $h^2\Omega_{\mathrm{exp}}(f)$ corresponds to the sensitivity of the experimental setup. 

For any future comparison purposes, we provide a benchmark point with input parameters: 
\begin{align}
m_\Sigma = 160 \; \text{GeV}, \quad \quad a_2 = 1.321,  \quad \quad b_4 = 1.0.
\end{align} 
This point admits a two-step phase transition, for which the second transition is of first order.
The parameters in the Lagrangian at the initial scale $\mu_0 = M_Z$ for this benchmark point possess the following values:
\begin{align}
&(y^2_t, g^2_s, g^2_2, g^2_1, \mu^2_{h}, \mu^2_\Sigma, \lambda_h, a_2, b_4)_{\text{tree-level}} \nonumber \\
&= (0.95386, 1.48409, 0.411406, 0.118004, 7812.5 \; \text{GeV}^2, 15898.1 \; \text{GeV}^2, 0.124351, 1.321, 1.0)  
\end{align} 
at tree-level, and
\begin{align}
&(y^2_t, g^2_s, g^2_2, g^2_1, \mu^2_{h}, \mu^2_\Sigma, \lambda_h, a_2, b_4)_{\text{1-loop}} \nonumber \\
&= (0.964377, 1.48409, 0.424624, 0.128051, 8050.76 \; \text{GeV}^2, 15507.4 \; \text{GeV}^2, 0.135361, 1.321, 1.0)   
\end{align} 
at one-loop corrections \cite{Niemi:2020hto,Niemi:2018asa}. 
We systematically evolve all these parameters to the scale $\mu = \pi T$ utilizing one-loop $\beta$-functions specific to the model. Subsequently, we employ the dimensional reduction method outlined in Refs.~\cite{Niemi:2020hto,Niemi:2018asa} to derive the parameters within the Lagrangian of the 3d EFT. In this context, we consider the triplet scalar field to be dynamic, and set the renormalization scale of the 3d EFT to be $\mu_{3\rm{d}} = T$.

For this specific benchmark point, we determine the critical temperature of the $\Sigma\rightarrow H$ transition to be $T_c = 108.54$ GeV, along with the following thermal parameters:
\begin{align}
T_* = 74.78\, {\rm GeV}, \quad \quad \alpha = 2.965 \times 10^{-2}, \quad \quad \frac{\beta}{H_*} = 17.37 \,. 
\end{align}
The SNR from the LISA detector, plotted as a function of wall velocity for this benchmark point, is depicted by the dashed black line in Fig.~\ref{fig:vw}.

\section{$\Sigma$SM phenomenology}
\label{sec:appendix-pheno}

Phase transition thermodynamics and possible resulting gravitational wave signatures depend sensitively on the scalar couplings, but these couplings are notoriously difficult to measure experimentally at colliders. This is in particular the case for the triplet self-coupling $b_4$. On the other hand, triplet-Higgs portal coupling $a_2$ has reasonable prospects of being experimentally measured. This could be achieved by precisely measuring the triplet loop contribution to the Higgs diphoton decay width, or by precisely measuring the branching fractions of the neutral triplet. We emphasize, that at one-loop level, both of these probes are independent of $b_4$. 

\begin{figure}[t]
    \begin{tabular}{c c}
    \includegraphics[width=0.45\textwidth]{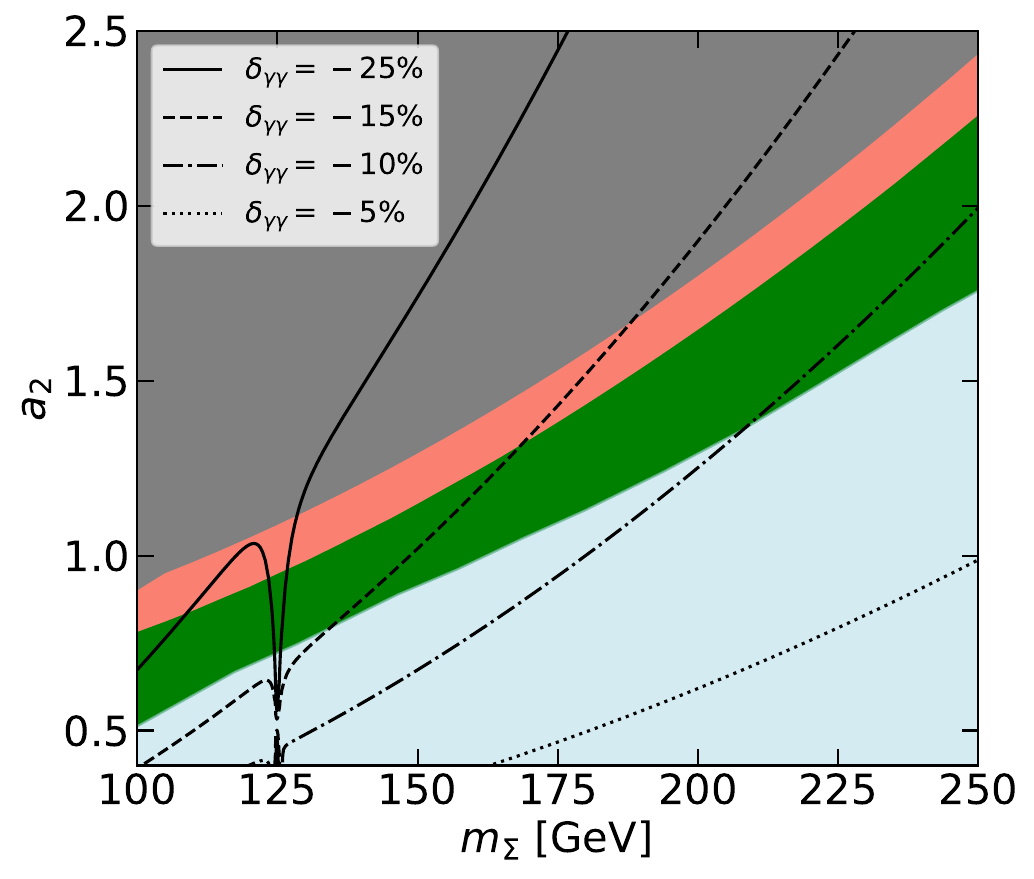}
    \includegraphics[width=0.45\textwidth]{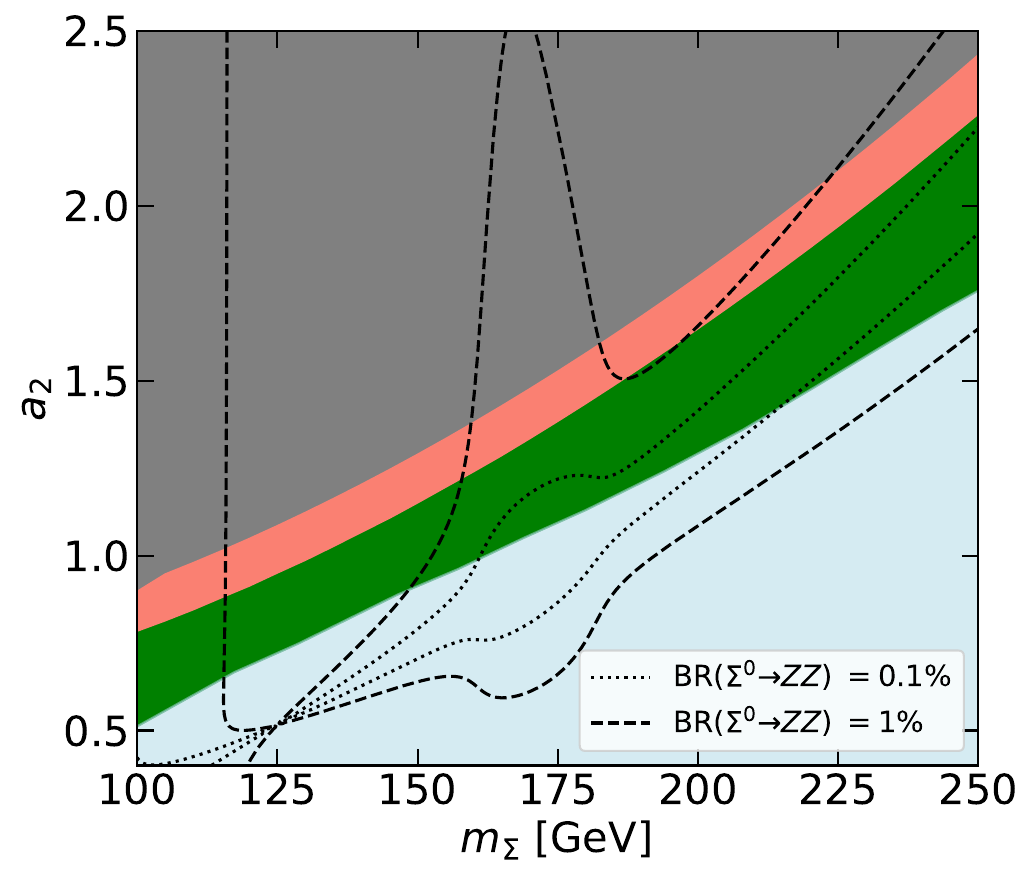}
  \end{tabular}
	\caption{
	Contours for phenomenological quantities of interest in $(m_\Sigma,a_2)$-plane for $\delta_{\gamma\gamma}$ (left) and $\mathrm{BR}(\Sigma^0 \rightarrow Z Z)$ (right), on top of thermodynamic phase structure of Fig.~\ref{fig:fig1}(a). For illustration, simplified expressions for both quantities are given in the main text. In both plots, we include small, non-zero triplet VEV $v_\Sigma = 1$ GeV. This generates mixing between Higgs and triplet and leads to singular behaviour for $\delta_{\gamma\gamma}$ around $m_\Sigma \sim m_H$, as well as makes neutral triplet unstable, which allows its decay to $ZZ$ and other channels.  
	}
	\label{fig:pheno}
\end{figure}

The dominant contributions to the Higgs diphoton decay width come from one-loop diagrams featuring the top quark and $W^\pm$ bosons. With the addition of the charged triplet scalar,
the Higgs diphoton decay width~\cite{Djouadi:2005gi,Inoue:2015pza} reads
\begin{equation}
\label{eq:diphoton}
\begin{aligned}
  \Gamma^{\mathrm{\Sigma SM}}_{h \rightarrow \gamma \gamma}
  = & 
    \frac{\alpha^2 g_2^2}{1024 \pi^3} \frac{m_{h}^3}{m_W^2}
     \left\lvert\
     \frac{4}{3} F_{1/2}\left(x_t\right)
     + F_1 \left(x_W\right) 
     + a_2 \frac{v_H^2}{ 2 m_{\Sigma}^2}
      F_{0}\left(x_\Sigma\right) \right\rvert ^2    \, ,
\end{aligned}
\end{equation}
where $x_i$ = $4 \frac{m_i^2}{m_{h}^2}$ and the $F_{0,1/2,1}$ are loop functions,
\begin{subequations}
  \begin{align}
    F_{0} (x) &= x \Big(1 - x f(x) \Big), \\
    F_{1/2} (x) &= - 2 x \Big(1 + (1 - x)f(x) \Big), \\
    F_{1} (x) &= 2 + 3 x \Big(1 + (2-x)f(x) \Big)\\
    f(x) & =
           \begin{cases} 
             \arcsin^2 (1/\sqrt{x}) & x\geq 1 \\
             - \frac{1}{4}\left( \ln\frac{1 + \sqrt{1-x}}{1-\sqrt{1-x}} - i \pi \right)^2 &  x < 1 \\
           \end{cases} \, .
  \end{align}
\end{subequations}
The SM result follows by taking $a_2 \rightarrow 0$ in Eq.~\eqref{eq:diphoton}. 
Measuring the deviation of the Higgs diphoton decay rate from the SM prediction will directly constrain $a_2$ and $m_\Sigma$. We quantify this deviation using,
\begin{equation}
    \delta_{\gamma \gamma} \equiv \frac{\Gamma^{\mathrm{\Sigma SM}}_{h \rightarrow \gamma \gamma}}{\Gamma^{\mathrm{SM}}_{h \rightarrow \gamma \gamma}} - 1.
\end{equation}
We show this quantity as a function of $(m_\Sigma, a_2)$ in the left panel of Fig.~\ref{fig:pheno}.
For simplicity, in Eq~\eqref{eq:diphoton} we have neglected possible mixing between the Higgs and triplet, that would be present in a case of non-zero triplet VEV $v_\Sigma$.
Such mixing only plays a significant role when $m_\Sigma \sim m_H$, and we demonstrate that in the aforementioned figure that is based on more general computation.

\begin{figure}[t]
\centering
\includegraphics[width=0.45\textwidth]{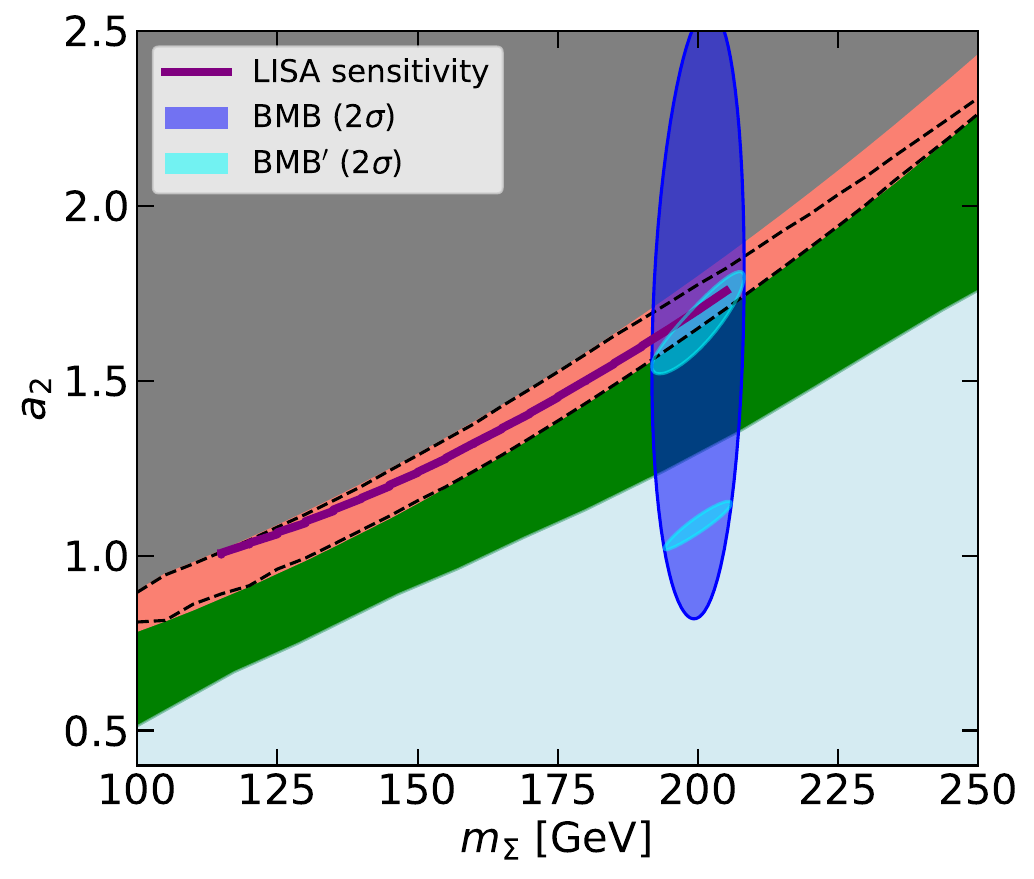}
\caption{
Same as Fig.~\ref{fig:fig1}(c), but the uncertainty on $\delta_{\gamma\gamma}$ is taken to be $\sigma_{\delta_{\gamma\gamma}}= 0.04$ which can be archived at the HL-LHC \cite{Cepeda:2019klc}. 
}
\label{fig:HL-LHC}
\end{figure}

\begin{figure}[t]
\centering
\includegraphics[width=0.45\textwidth]{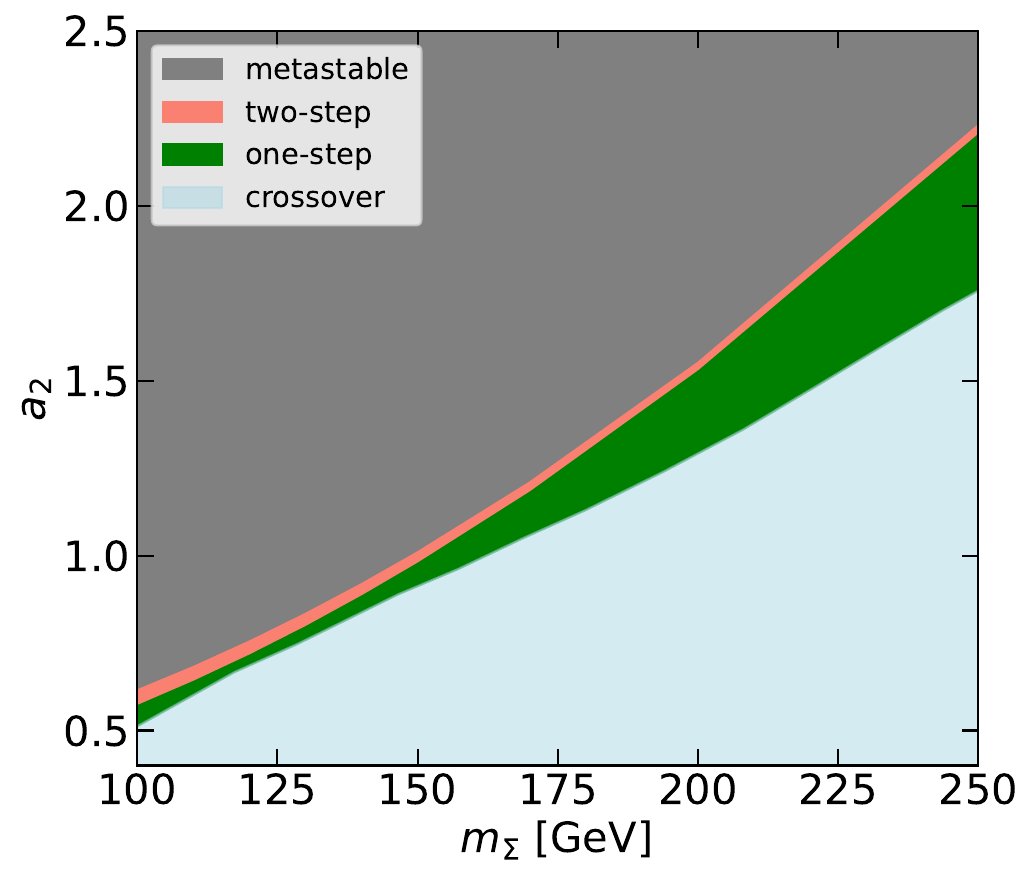}
\caption{
Same as Fig.~\ref{fig:fig1}(a) but for $b_4 = 0.25$.
Relative to Fig.~\ref{fig:fig1}(a) two-step region is significantly
narrower.
}
\label{fig:vary-b4}
\end{figure}

Another means to determine $a_2$ (in non-$Z_2$ symmetric case) is by measuring the neutral triplet's branching ratios. For low triplet mass, the triplet will decay primarily into pairs of fermions via its mixing with the Higgs boson. For large triplet mass ($\gtrsim 120$ GeV) the neutral triplet will decay primarily into pairs of bosons ($W^\pm W^{\mp}$, $Z Z$, $h h$). 
These decays depend sensitively on $a_2$, and in particular
when \cite{Bell:2020gug}
\begin{equation}
   \mu_\Sigma^2 = v^2_\Sigma b_4,
\end{equation}
the off-diagonal terms in the neutral scalar mass matrix disappear, resulting in
\begin{equation}
    \Gamma_{\Sigma^0 \rightarrow Z Z} \ =\  \Gamma_{\Sigma^0 \rightarrow h h}\  =\ 0, \quad \Gamma_{\Sigma^0 \rightarrow W W} \ \approx \ 1.
\end{equation}
The region of parameter space where this cancellation occurs is
close to
the region where the two-step phase transition takes place. 
For this reason,
the branching ratios are 
sensitive to the value of $a_2$ in the two-step region,
providing a mean to measure it.
Since, the $\Sigma^0 \rightarrow h h^*$ branching ratio strongly suppressed when $m_\Sigma < 2 m_h$, we will focus on the $Z Z$ decay. When 
$m_\Sigma > 2 m_Z$,
the branching ratio for this decay can be 
approximated using the tree level two-body partial widths, resulting
\begin{align}
    \mathrm{BR}(\Sigma^0 \rightarrow Z Z) &\approx \frac{\Gamma_{\Sigma^0 \rightarrow Z Z}}{\Gamma_{\Sigma^0 \rightarrow W W}}
     = \frac{ \sqrt{m_{\Sigma}^2 - 4 m_W^2} \left(m_{\Sigma}^4 - 4 m_{\Sigma}^2 m_W^2 + 12 m_W^4 \right)}
    {\frac{1}{2}\sqrt{m_{\Sigma}^2 - 4 m_Z^2}  \left(m_{\Sigma}^4 - 4 m_{\Sigma}^2 m_Z^2 + 12 m_Z^4 \right)} \left(\frac{4 v_\Sigma}{v_H} \cot \theta + 1 \right)^2 .
\end{align}
This expression is defined in the mass eigenstate basis 
\begin{align}
  \begin{pmatrix}
    h_1 \\ h_2 
  \end{pmatrix}
  &= 
  \begin{pmatrix}
    \cos \theta &\   - \sin \theta \\
    \sin \theta &\  \cos \theta
  \end{pmatrix} 
  \begin{pmatrix}
    H^0 \\ \Sigma^0 
  \end{pmatrix},
\end{align}
where mixing angle $\theta$ is defined to
diagonalise neutral scalar sector
\begin{equation}
\begin{aligned}
  \frac{1}{2}\begin{pmatrix}
    H^0 & \Sigma^0 
  \end{pmatrix}
  \mathcal{M}_{N}
  \begin{pmatrix}
    H^0 \\ \Sigma^0 
  \end{pmatrix}
   = \frac{1}{2}\begin{pmatrix}
    h_1 & h_2 
  \end{pmatrix}
  \begin{pmatrix}
    m_{h}^2 & 0 \\ 0 & m_{\Sigma}^2
  \end{pmatrix}
  \begin{pmatrix}
    h_1 \\ h_2 
  \end{pmatrix}, 
\end{aligned}
\end{equation}
where $m_{h} = 125.1$ GeV is the observed Higgs boson mass. 
The neutral scalar mass matrix reads
\begin{align}
\label{eq:massMatrix}
  \mathcal{M}_N &= 
  \begin{pmatrix}
    2 \lambda_H v_H^2  \  &\  a_2 v_H v_\Sigma - \frac{1}{2} a_1 v_H
    \\
    a_2 v_H v_\Sigma - \frac{1}{2}a_1 v_H\ & \ - \mu^2_\Sigma + \frac{1}{2} a_2 v_H^2+  3 b_4 v_\Sigma^2   
 \end{pmatrix}.
\end{align}
The $a_1$ term in the potential and the
triplet's non-zero VEV 
are responsible of
mixing between the neutral component of the
triplet and SM Higgs.
In right panel of Fig.~\ref{fig:pheno} we visualise contours of $\mathrm{BR}(\Sigma^0 \rightarrow Z Z)$ near two-step region. 
A non-vanishing $v_\Sigma$ ensures that the neutral component of the triplet unstable, which is also required to avoid 
dark matter direct detection constraints~\cite{Bell:2020gug, Katayose:2021mew}. However, the VEV has to be sufficiently small to satisfy constraints from electroweak precision observables, which require~\cite{Zyla:2020zbs}
\begin{align}
& \rho_0 = 1 + \frac{4 v_\Sigma^2}{v_H^2} = 1.0004 \pm 0.0002 \,,
\end{align}
which implies $v_\Sigma \lesssim 3.5$ GeV at $2\sigma$ accuracy.

To obtain 2$\sigma$ favored regions from the collider measurements BMA$'$ in the main body, 
we define
\begin{equation}
    \chi^2 = \frac{\left({\rm BR}{(\Sigma^0 \to ZZ)}^{\rm th} - {\rm BR}{(\Sigma^0 \to ZZ)}^{\rm obs}\right)^2}{\sigma_{{\rm BR}({\Sigma^0 \to ZZ})}^2} 
    + \frac{\left(\delta_{\gamma \gamma}^{\rm th} - \delta_{\gamma \gamma}^{\rm obs}\right)^2}{\sigma_{\delta_{\gamma \gamma}}^2} 
    + \frac{(m_\Sigma^{\rm th} - m_\Sigma^{\rm obs})^2}{\sigma_{m_\Sigma}^2},
\end{equation}
where $\sigma$ indicates the uncertainty, and superscripts $th$ and $obs$ represent theory and observed quantities respectively. We note that, in this analysis, the collider measurements of $m_\Sigma$, ${\rm BR}({\Sigma^0 \to ZZ})$ and $\delta_{\gamma \gamma}$ are hypothetical measurements. 
We then demand $\chi^2 < 2.71$ to obtain 2$\sigma$ favoured region. 

In Fig.~\ref{fig:HL-LHC} we present a plot in analogy to Fig.~\ref{fig:fig1}(c).
We take the uncertainty on $\delta_{\gamma \gamma}$ from HL-LHC \cite{Cepeda:2019klc}, 
\begin{align}
\delta_{\gamma\gamma} =  -0.132 \pm 0.04, 
\end{align}
while the triplet mass and ${\rm BR}{(\Sigma^0 \to ZZ)}$ measurements are kept the same as in BMA. 
We denote this benchmark without (with) $\text{BR}(\Sigma^0 \to ZZ)$ measurement as  BMB (BMB$'$).
   
Compared to BMA, the worse accuracy reflects to much wider $2\sigma$ region (blue) in Fig.~\ref{fig:HL-LHC}, making such measurement alone indecisive for character of the phase transition, as blue region spans from two-step to one-step to crossover. When accompanied with measurement of $\mathrm{BR}(\Sigma^0 \rightarrow Z Z)$ (BMB$'$) corresponding $2\sigma$ region overlaps with two-step region and LISA sensitivity, but note that there is another possible, disconnected region in crossover regime, implying that neither BMB$'$ alone can be decisive to determine character of the EWPT. 

While $\delta_{\gamma\gamma}$ and $ \mathrm{BR}(\Sigma^0 \rightarrow Z Z)$ are independent on $b_4$ at leading order, thermodynamic phase structure depends on 
this parameter.
 To illustrate this, in Fig.~\ref{fig:vary-b4} we show a similar plot as Fig.~\ref{fig:fig1}(a) but fixing $b_4 = 0.25$. Comparing with Fig.~\ref{fig:fig1}(a), one observes that two-step region is significantly narrower for a smaller value of $b_4$; requiring a degree of fine-tuning for a model to live in this region. 
Similar observation has been made e.g. in~\cite{Kurup:2017dzf} in case of the real singlet extension model.
We note that Fig.~\ref{fig:vary-b4} reproduces 
Fig. 1 in \cite{Niemi:2020hto}, except that here we are unable to delineate the
boundary between the first-order and the crossover regions in two step transitions.

\begin{figure*}[t]
  \includegraphics[width=0.45\textwidth]{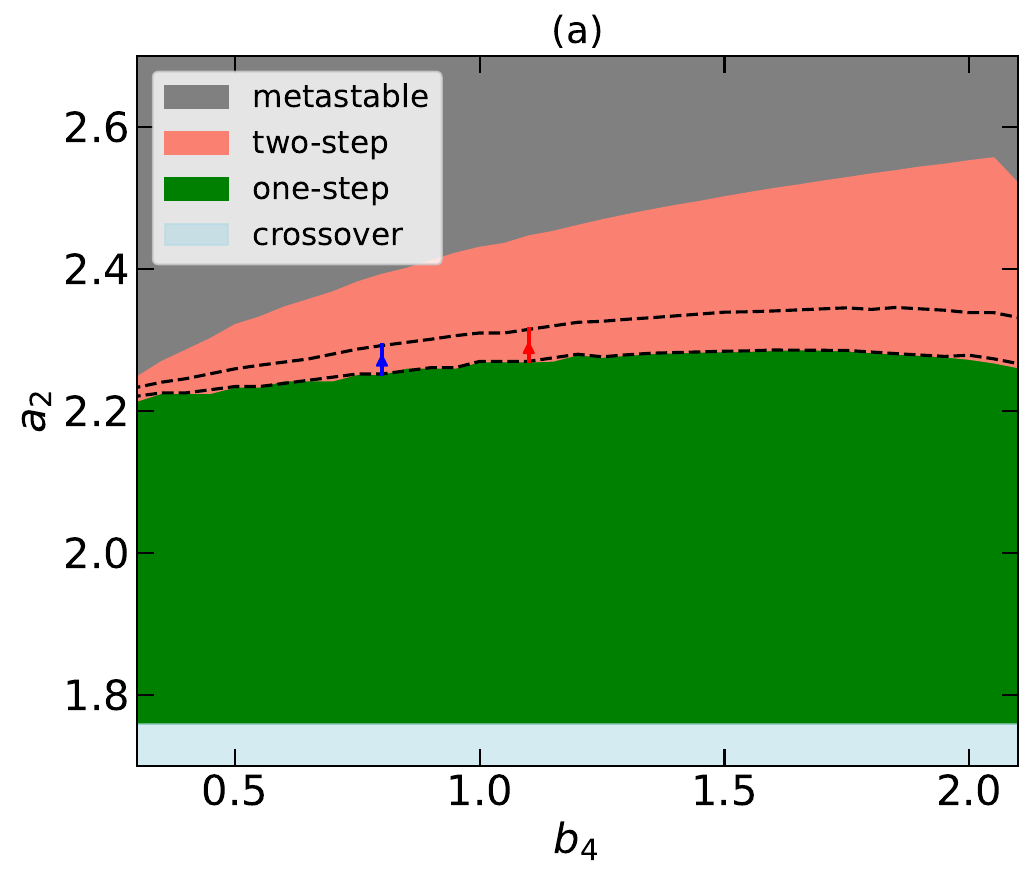} 
  \includegraphics[width=0.45\textwidth]{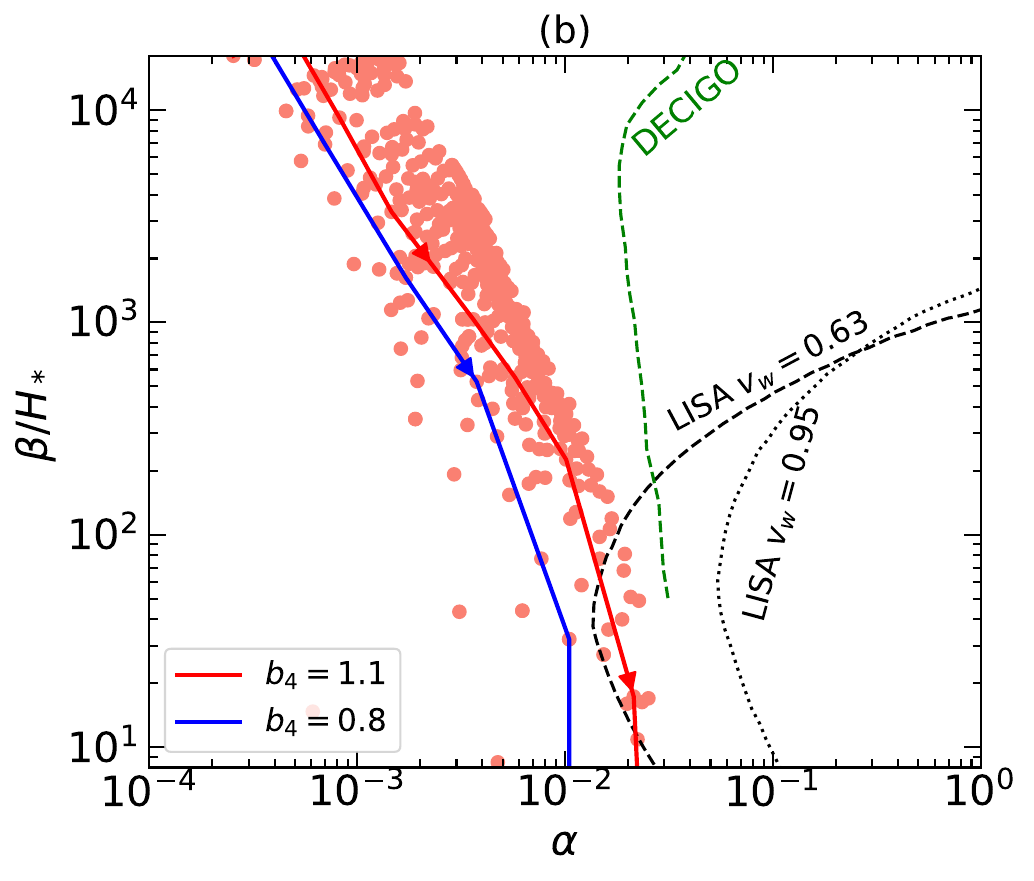} 
  \includegraphics[width=0.45\textwidth]{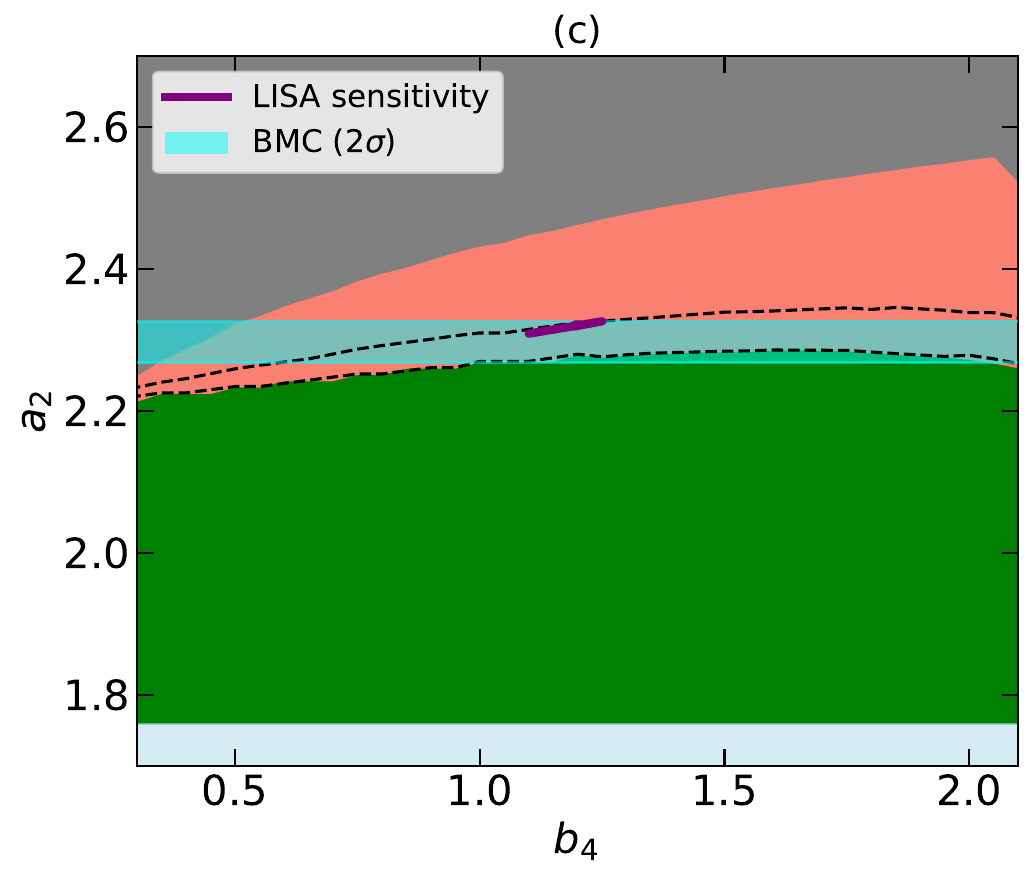} 
  \includegraphics[width=0.45\textwidth]{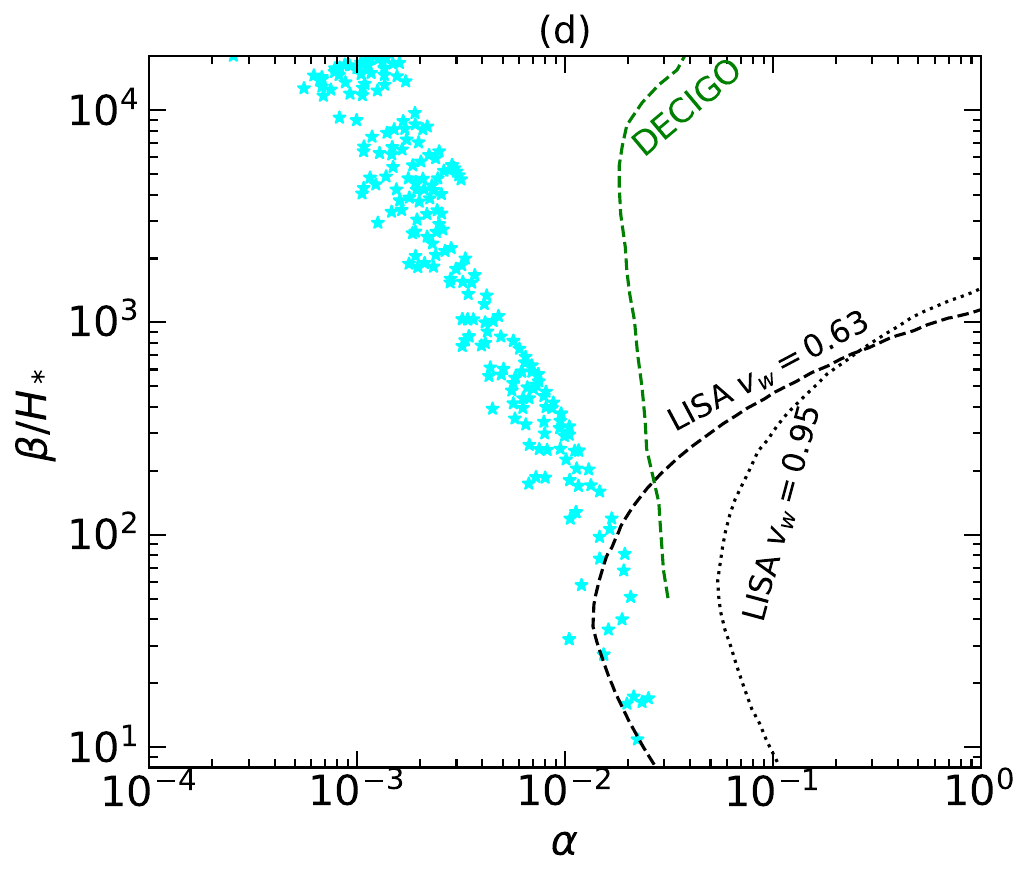} 
  \caption{Same as Fig.~\ref{fig:fig1} but in ($b_4, a_2$) plane with fixed $m_\Sigma = 250$ GeV. In panels (a) and (b), the blue and red lines correspond to $b_4 = 0.8$ and $1.1$ respectively. The cyan band in panel (c) and blue star points in panel (d) represent the $2\sigma$ favored region from BMC measurements.} 
  \label{fig:fig4}
\end{figure*}

For further illustration, we present yet another choice for a benchmark point of the main body. We define a benchmark set BMC: 
\begin{align}
\text{BR}(\Sigma^0 \to ZZ) = 0.002 \pm 0.0004, 
\end{align}
with the value for $\delta_{\gamma\gamma}$ taken to be the same as in BMA. The central value of the $\text{BR}(\Sigma^0 \to ZZ)$ corresponds to $m_\Sigma = 250$ GeV and $a_2 = 2.3$, and the uncertainty is taken to be $20\%$ of the central value.

Fig.~\ref{fig:fig4} presents an analysis similar to that of Fig.~\ref{fig:fig1}, but with key distinctions: we explore the ($b_4, a_2$) plane with the triplet mass $m_\Sigma$ held constant at 250 GeV, treating $m_\Sigma$ as a known parameter.

In Fig.~\ref{fig:fig4}(a), we examine how the thermodynamic phase structure varies as a function of $b_4$. Our findings indicate that the nucleation region for the two-step transition (enclosed by the dashed black line) exhibits a mild dependence on $b_4$. The thermal parameters, $\alpha$ and $\beta/H_{*}$, show a significant sensitivity to $a_2$, as illustrated in Figure~\ref{fig:fig4}(b). Consistent with the observations in Fig.~\ref{fig:fig1}(b), an increase in $a_2$ leads to a higher $\alpha$ and a lower $\beta/H_{*}$. Furthermore, a decrease in $b_4$ value results in a stronger phase transition, thereby moving into regions that future GW detectors will be able to probe.

The 2$\sigma$ preferred regions for BMC are highlighted by a cyan band and cyan star points in Figs.~\ref{fig:fig4}(c) and (d), respectively.
The lesson we illustrate is, that if
1) the triplet mass is known from collider experiments,  2) LISA discovers GW background, and 3) we assume that such a background is a remnant of a strong first order phase transition in the triplet model, we get relatively stringent constraint on $a_2$ (as well as lower bound --  and relatively less tight constraint -- on $b_4$) that translates roughly to collider measurements of BMC.
On the other hand, a null result of LISA is still viable together with measurement BMC; this puts on upper bound on $b_4$ (as function of $a_2$).

\end{document}